\pdfoutput=1

\documentclass[11pt]{article}

\usepackage[final]{acl}

\usepackage{times}
\usepackage{latexsym}

\usepackage[T1]{fontenc}

\usepackage[utf8]{inputenc}

\usepackage{microtype}

\usepackage{inconsolata}

\usepackage{graphicx}

\usepackage{mdframed}
\usepackage{listings}
\usepackage{xcolor}
\usepackage{array}
\usepackage{booktabs}
\usepackage{multirow}
\usepackage{amsmath}
\usepackage{colortbl}
\usepackage{pifont}
\usepackage{makecell}

\usepackage[ruled,vlined]{algorithm2e}

\definecolor{codebg}{rgb}{0.99,0.99,0.99}
\definecolor{highlightyellow}{rgb}{1,1,0.8}
\definecolor{highlightpink}{rgb}{1,0.8,0.8}
\definecolor{highlightgreen}{rgb}{0.2,0.85,0.2}
\definecolor{darkblue}{rgb}{0.0, 0.0, 0.55}

\definecolor{keyword}{RGB}{0, 0, 255}
\definecolor{comment}{RGB}{0, 128, 0}
\definecolor{string}{RGB}{163, 21, 21}
\definecolor{background}{RGB}{255, 255, 255}
\definecolor{identifier}{RGB}{128, 0, 128}

\newcommand{\lstcolourline}[1]{\rlap{\color{#1}\rule[-.3\baselineskip]{7.35cm}{\baselineskip}}}
\definecolor{mycolor}{rgb}{0.95,0.95,0.95}

\newmdenv[
  backgroundcolor=mycolor,
  linecolor=black,
  linewidth=1pt,
  roundcorner=10pt,
  innerleftmargin=10pt,
  innerrightmargin=10pt,
  innertopmargin=10pt,
  innerbottommargin=10pt
]{custombox}

\lstdefinestyle{mystyle}{
    language=Python,
    basicstyle=\ttfamily\small,
    keywordstyle=\color{blue},
    stringstyle=\color{highlightgreen},
    commentstyle=\color{purple},
    backgroundcolor=\color{codebg},
    numbers=left,
    numberstyle=\footnotesize\ttfamily\hfill, 
    numbersep=3pt, 
    frame=tb,
    rulecolor=\color{black},
    framerule=0.5mm,
    framesep=2mm,
    xleftmargin=4.5mm,
}

\lstdefinestyle{customlst}{
    frame=topline,
    framexbottommargin=5pt,
    framextopmargin=5pt,
    xleftmargin=5pt,
    xrightmargin=5pt,
    aboveskip=10pt,
    belowskip=10pt,
    language=Python,
    basicstyle=\ttfamily,
    keywordstyle=\color{blue}\bfseries,
    commentstyle=\color{gray},
    stringstyle=\color{red},
    morekeywords={Line Number, Type, Explanation, Dead code},
    breaklines=true,
    tabsize=2,
    numbers=left,
    numberstyle=\tiny\color{gray},
    stepnumber=1,
    numbersep=5pt,
    showspaces=false,
    showstringspaces=false
}

\lstdefinestyle{JavaStyle}{
    backgroundcolor=\color{background},
    commentstyle=\color{comment},
    keywordstyle=\color{keyword},
    stringstyle=\color{string},
    basicstyle=\ttfamily\footnotesize,
    breakatwhitespace=false,
    breaklines=true,
    captionpos=b,
    keepspaces=true,
    numbers=none,
    showspaces=false,
    showstringspaces=false,
    showtabs=false,
    tabsize=2,
    language=Java
}

\lstdefinestyle{PythonStyle}{
    keywordstyle=\color{blue},
    stringstyle=\color{highlightgreen},
    backgroundcolor=\color{background},
    commentstyle=\color{comment},
    basicstyle=\ttfamily\footnotesize,
    breakatwhitespace=false,
    breaklines=true,
    captionpos=b,
    keepspaces=true,
    numbers=none,
    showspaces=false,
    showstringspaces=false,
    showtabs=false,
    tabsize=2,
    language=Python
}

\title{DCE-LLM: Dead Code Elimination with Large Language Models}

\author{
 Minyu~Chen$^{1}$, Guoqiang Li$^{1}\thanks{~~Corresponding author.}$ , Ling-I Wu$^{1}$, Ruibang Liu$^{1}$ \\
 $^{1}$Shanghai Jiao Tong University, Shanghai, China\\
 \texttt{{\{minkow,li.g,edithwuly,628628\}@sjtu.edu.cn}}
}

\begin{document}
\maketitle
\begin{abstract}
Dead code introduces several challenges in software development, such as increased binary size and maintenance difficulties. It can also obscure logical errors and be exploited for obfuscation in malware. For LLM-based code-related tasks, dead code introduces vulnerabilities that can mislead these models, raising security concerns.
Although modern compilers and IDEs offer dead code elimination, sophisticated patterns can bypass these tools. A universal approach that includes classification, location, explanation, and correction is needed, yet current tools often require significant manual effort.
We present DCE-LLM, a framework for automated dead code elimination using a small CodeBERT model with an attribution-based line selector to efficiently locate suspect code. LLMs then generate judgments and explanations, fine-tuned on a large-scale, annotated dead code dataset to provide detailed explanations and patches. DCE-LLM outperforms existing tools, with advanced unreachability detection, automated correction, and support for multiple programming languages. Experimental results show DCE-LLM achieves over 94\% F1 scores for unused and unreachable code, significantly surpassing GPT-4o by 30\%.\footnote{Our dataset and code are available at \url{ https://github.com/Minkow/DCE-LLM}}
\end{abstract}

\section{Introduction}

Dead code, defined as segments of a program that are never executed or that do not affect program functionality, can introduce performance bottlenecks and obscure potential security vulnerabilities. The presence of dead code increases binary size, which negatively impacts system performance, particularly in resource-constrained environments like web applications~\cite{8370748,malavolta2023javascript}. Additionally, dead code complicates software maintenance by cluttering the codebase, thus consuming developers' time as they attempt to comprehend its purpose. In more critical cases, dead code can serve as a vector for obfuscation techniques employed in adversarial attacks, complicating code comprehension and analysis.

Though large language models (LLMs) have achieved impressive results in various code-related tasks~\cite{weicoeditor, jainllm, Tang2024CodeAgentAC, li2024rewriting, han2024archcode}, they are particularly vulnerable when faced with dead code. Dead code, whether introduced unintentionally or as part of an attack, can disrupt a model's ability to accurately interpret program logic. In particular, dead code injection attacks exploit this vulnerability by adding redundant or unreachable code, which confuses the model and leads to incorrect predictions or faulty analyses. Studies have shown that this can cause a significant drop in accuracy for tasks like vulnerability detection, with an observed reduction of up to 12.7\%~\cite{khare2023understanding}. These weaknesses highlight the need for more robust approaches to handling dead code in LLM-based tasks.

While modern compilers can remove basic dead variables and IDEs can flag some dead code~\cite{wang2017dead, romano2016graph}, effectively eliminating all forms of dead code remains a challenge. Moreover, sophisticated attack patterns using unreachable statements can bypass these tools, leading to potential security vulnerabilities and performance issues. These attacks exploit weaknesses in language models, embedding dead code that remains undetected. Additionally, the pervasiveness of dead code across different programming languages complicates the development of a universal solution, as each language has its own syntax and semantics. As a result, a robust and versatile dead code elimination framework is essential, one that can automate the processes of classification, detection, explanation, and correction, thereby reducing the need for manual intervention.

LLMs face several key challenges when it comes to dead code elimination. First, LLMs are trained on publicly available code datasets, many of which contain dead code. However, these datasets lack specific labels for dead code, meaning LLMs do not explicitly learn to identify or handle it during training. For example, we find that over 45.3\% of Java code in the CodeNet~\cite{puri2021codenet} dataset contains dead code, but this is not filtered or annotated, leading to gaps in the models' understanding.
Second, LLMs struggle to locate dead code in long inputs. Dead code can appear in various parts of a program, often scattered across lines of code. LLMs have difficulty maintaining robust context over long inputs, making it challenging to accurately detect and address dead code in complex or extended codebases.
Third, while LLMs excel at tasks like code generation and refactoring, explaining and correcting dead code presents a different challenge. Properly addressing dead code requires a deep understanding of the intended functionality and context of the program, which LLMs may fail to grasp. Additionally, generating semantically correct patches for dead code requires advanced reasoning capabilities that go beyond simple code completion or modification tasks.

To address these challenges, we present DCE-LLM, an LLM-empowered framework for comprehensive dead code elimination. We leverage a relatively small CodeBERT model with a novel attribution-value-based line selector to effectively and accurately locate suspect dead code snippets in long code inputs. This step filters out most normal code, reducing the call time for LLMs. Subsequently, LLMs generate final judgments and explanations for dead code, focusing on the highlighted suspect lines. Furthermore, we fine-tune LLMs on the first large-scale dead code dataset, which is automatically annotated, equipping them with the capability to generate explanations and patches for dead code. As the first neural-based framework for dead code elimination, DCE-LLM demonstrates several advantages over existing tools:

\textit{Unreachability Checking.} DCE-LLM effectively detects complex unreachable dead code that can bypass compilers and IDEs, addressing vulnerabilities used in adversarial attacks.

\textit{Explanation and Correction.} DCE-LLM automatically provides explanations and patches for dead code, significantly reducing the manual effort required by developers and enhancing overall development efficiency.

\textit{Programming Language Support.} DCE-LLM leverages the natural ability of LLMs to understand and generate code in multiple programming languages. It supports multiple programming languages, including Python and Java, making it applicable across diverse development environments.

Our experimental results demonstrate that DCE-LLM achieves F1 scores over 94\% for both unused and unreachable code, along with F1 scores significantly surpassing those of existing LLMs and IDEs. Moreover, in terms of the quality of generated content, including explanations and repaired code, DCE-LLM outperformed baseline models in arena-style human evaluations.

We highlight our contributions as follows:
\begin{itemize}
\item We propose a novel framework, DCE-LLM, for leveraging LLMs to tackle the dead code elimination task, encompassing classification, location, exploration, and patching. To our knowledge, this is the first application of LLMs for dead code elimination, offering sophisticated unreachability checking, detailed explanations, and effective code repair.
\item We employ a small CodeBERT model with an innovative attribution technique to accurately locate suspect lines, augmenting LLMs with extra those hints.
\item We introduce the first large-scale dead code dataset, demonstrating that DCE-LLM achieves over 94\% F1 scores in detecting unused and unreachable code across multiple programming languages, surpassing GPT-4o with over 30\% F1.
\end{itemize}

\section{Related Work}
\textbf{Dead Code Elimination}. Researchers have proposed various approaches for detecting dead codes. Chen et al.~\cite{chen1998c++} introduced a data model serving reachability
analysis and dead code detection for C++ repositories. Boomsma et al.~\cite{boomsma2012dead} leverages a dynamic framework for extracting dead files in PHP web apps by monitoring file usage. Several works focus on call graph analysis~\cite{romano2016graph, romano2018exploring} while program slicing is also adopted to build generic frameworks~\cite{alabwaini2018using, wang2017dead}
. Very recently, Lacuna~\cite{malavolta2023javascript} has presented to integrate third-party analysis techniques for JavaScript dead code detection.

Our proposed DCE-LLM takes a fundamentally different approach.  As a learning-based method, it avoids reliance on specific programming languages or external code analysis tools like LLVM~\cite{wang2017dead}, or hand-crafted rules. This allows DCE-LLM to generalize across multiple languages and handle incomplete or even non-compilable code, making it both versatile and robust.  Critically, our method effectively addresses the challenge of adversarial unreachable code and offers practical solutions for dead code repair, completing the dead code elimination pipeline rather than just performing detection.


\textbf{LLMs for Code.} LLMs have demonstrated remarkable capabilities in various code-related tasks beyond generation and completion. Recent research highlights their effectiveness in areas like bug detection~\cite{wang-etal-2024-sanitizing}, code transpilation~\cite{bhatia2024verified}, and code repair~\cite{tang2024code, zhao-etal-2024-repair}. For example, NS-Slicer~\cite{yadavally2024learning} fine-tuned an LLM to predict static program slices for both complete and partial code, subsequently using these predictions to enhance vulnerability detection. These advancements showcase the code reasoning abilities of LLMs in diverse scenarios and provide a strong foundation for the effectiveness of DCE-LLM.

\section{Background}

\begin{figure}[!bp]
\begin{minipage}{\linewidth}
\small
\begin{lstlisting}[style=mystyle, escapechar=!]
def fill_str(Data):
  s1 = input()
  s2 = s1 + '<PAD>'
!\lstcolourline{yellow!25}!  s3 = s1 + '<EOS>'  # Unused Variable
!\lstcolourline{pink!20}!  if len(s2) == 0: # Unreachable Code
    print('Empty string')
    Data.pad_str = None
    Data.eos_str = None
  else:
    Data.pad_str = s2
    Data.eos_str = 's3'
\end{lstlisting}
\end{minipage}
\caption{An illustrative Python method with dead code.}
\label{fig1}
\end{figure}

As illustrated in Figure \ref{fig1}, Python method \texttt{fill\_str} contains several issues about two primary types of dead code. \textbf{Unused code} refers to code defined or executed but whose result is never used in any other computation. The execution of dead code wastes computation time and memory. In our example, The author of this method has accidentally put quotes around \texttt{s3} in line 11, resulting in an unused variable \texttt{s3} in the fourth line in light yellow.
\textbf{Unreachable code}, defined as code that can never be executed. Line 5 in light pink is the start of an unreachable code snippet. It introduces an unreachable case caused by an always-false condition. Thus, code in lines 6-8 cannot be executed whatever user input in line 2.

Modern compilers and IDEs offer extensive support for dead code elimination, especially for unused code. However, detecting unreachable code is inherently challenging, especially those designed for obfuscation or adversarial attacks as illustrated in Figure \ref{fig1}.
Static analysis tools might not catch this unreachable code \texttt{if len(s2) == 0:} if the condition requires to be evaluated with a satisfiability check. Moreover, several attack patterns are introduced~\cite{gao2023discrete} for producing complex unreachable code branches which are imperceptible to compilers and IDEs. They successfully mislead the output result of language models such as CodeBERT. Even powerful LLMs like GPT-4o are occasionally affected, without realizing that the dead branch is unreachable in the vulnerability detection task~\cite{khare2023understanding}.

\section{Methodology}

\begin{figure*}[!ht]
\centering
\includegraphics[width=16cm]{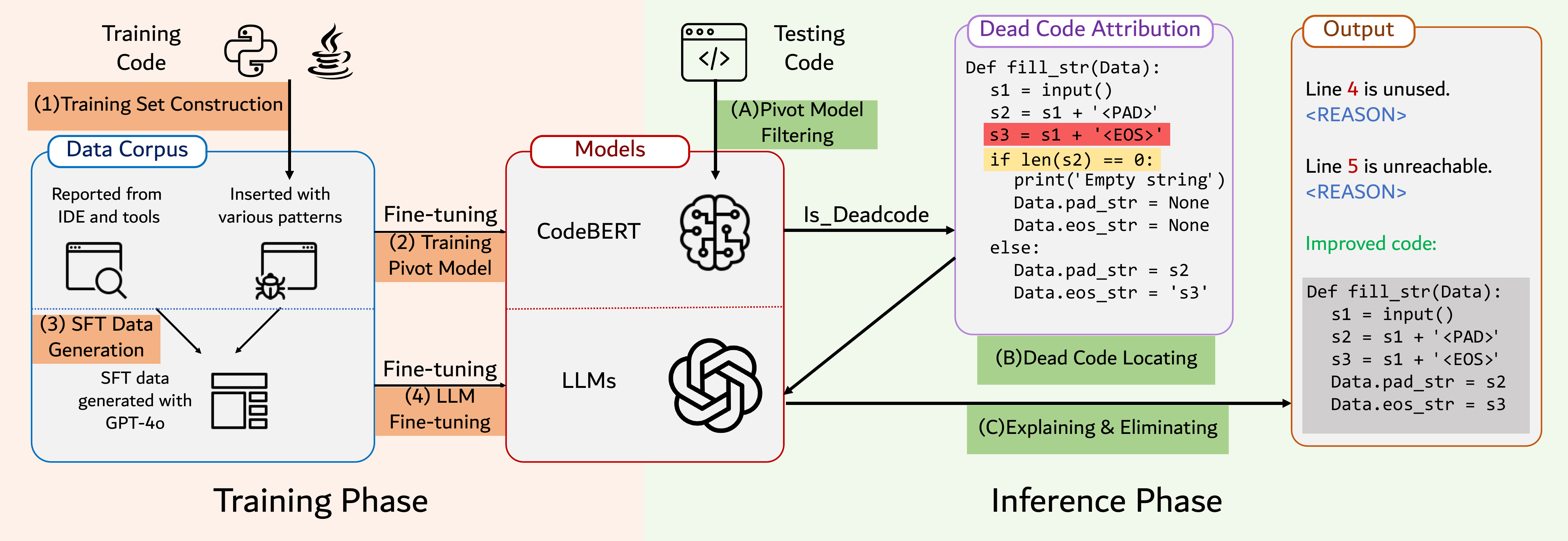}
\caption{Overview of the DCE-LLM approach. The yellow labels highlight the training step while the green labels introduce the inference step.}
\label{fig3}
\end{figure*}

This section introduces DCE-LLM (Figure \ref{fig3}), our novel approach for automated dead code elimination, comprising training (yellow) and inference (green) phases.  In training, we first train a CodeBERT-based pivot model for high recall on a constructed dead code dataset.  We then leverage GPT-4o to generate high-quality annotations, including explanations and improved code, for fine-tuning an LLM to enhance detection accuracy and code suggestion quality.  Inference begins with the pivot model filtering suspect code.  Our novel dead code attribution technique then precisely locates and incorporates potential dead code lines into LLM prompts. Finally, the LLM uses pivot model predictions as hints to generate practical dead code removal suggestions.

\subsection{Training Phase}

\subsubsection{Data Collection}
Systematically gathering a comprehensive dataset of both normal and dead code snippets is essential for our approach. This enables the model to learn the intricate details of code structure and functionality, improving its ability to detect and eliminate dead code precisely.

Unluckily, there are no large-scale datasets available specifically for dead code. Thus, we create a novel dataset, AIDCE, aimed at providing a comprehensive collection of annotated both unused and unreachable dead code samples.

In practice, we selected CodeNet~\cite{puri2021codenet} as our data source. CodeNet collects data from online judge websites, consisting of submissions to various programming problems. This dataset offers a rich collection of code snippets in various languages including C++, Python, and Java.
Since no code review process is conducted on online judge platforms, the code submissions often contain more dead code fragments written by programmers. From this corpus, we select code snippets written in Python and Java, resulting in about 300K code files.


We then leverage existing tools to annotate \textbf{unused} code. For Python, we use \textit{Vulture}~\cite{Vulture}, an open-source static code analyzer that detects unused functions, imports, and variables. Vulture can also identify unreachable code, but its capabilities are limited to code following a return statement and simple conditions tested with the \textit{eval()} method in Python. For Java, we utilize \textit{IntelliJ IDEA} from JetBrains~\cite{intellij2011most}. \textit{IntelliJ IDEA} offers a command-line inspector that operates in the background to perform inspections and highlight unused code. By harnessing the power of static analysis, we treat the unused code identified by these tools as gold-standard labels for our dataset.


\begin{table}[tbph]
    \centering
    \renewcommand{\arraystretch}{0.8}
    \small
    \begin{tabular}{>{\raggedright\arraybackslash}p{1cm}lp{0.8cm}p{0.8cm}}
        \toprule
        \multicolumn{1}{c}{\textbf{Name}} & \multicolumn{1}{c}{\textbf{Code}} & \textbf{Python} & \textbf{Java} \\
        \midrule
After return & {\begin{lstlisting}[style=PythonStyle]
return
{deadcode}
\end{lstlisting}}& {\textcolor{brown}{\ding{51}}} {\textcolor{darkblue}{\ding{51}}} & {\textcolor{brown}{\ding{51}}} {\textcolor{darkblue}{\ding{51}}}  \\
\midrule

        Covered branch & {\begin{lstlisting}[style=PythonStyle]
if a>b: {...}
elif a<=b: {...}
else: {deadcode}
\end{lstlisting}}& {\textcolor{brown}{\ding{55}}} {\textcolor{darkblue}{\ding{55}}} & {\textcolor{brown}{\ding{55}}} {\textcolor{darkblue}{\ding{51}}}  \\
\midrule

        Floor & {\begin{lstlisting}[style=PythonStyle]
b = math.floor(a)
if a<b: {deadcode}
\end{lstlisting}}& {\textcolor{brown}{\ding{55}}} {\textcolor{darkblue}{\ding{55}}} & {\textcolor{brown}{\ding{55}}} {\textcolor{darkblue}{\ding{55}}}  \\
\midrule

After assert & {\begin{lstlisting}[style=PythonStyle]
assert a>0
if a<0: {deadcode}
\end{lstlisting}}& {\textcolor{brown}{\ding{55}}} {\textcolor{darkblue}{\ding{55}}} & {\textcolor{brown}{\ding{55}}} {\textcolor{darkblue}{\ding{55}}}  \\
\midrule

Sorted array & {\begin{lstlisting}[style=PythonStyle]
a = sorted([...])
if a[0]>a[-1]:
    {deadcode}

\end{lstlisting}}& {\textcolor{brown}{\ding{55}}} {\textcolor{darkblue}{\ding{55}}} & {\textcolor{brown}{\ding{55}}} {\textcolor{darkblue}{\ding{55}}}  \\

\midrule
\multicolumn{4}{c}{ $\cdots$}\\

\midrule
\midrule
\multicolumn{4}{c}{\textcolor{brown}{Vulture: 1/62} \ \textcolor{darkblue}{PyCharm: 1/62} \ \ \  \textcolor{brown}{JIT: 1/62} \ \textcolor{darkblue}{IDEA: 18/62}}\\

\bottomrule
\end{tabular}
\caption{Examples of unreachable patterns}
\label{taba}
\end{table}

However, static analysis tools demonstrate vulnerabilities when encountering unreachable branch insertion attacks, as shown in Table \ref{taba}. These branches or loops contain conditions that always evaluate to false, yet are often complex enough to evade detection by compilers and analysis tools. In addition to attack patterns that deceive nearly all compilers and checkers, as illustrated by DaK~\cite{gao2023discrete}, we have expanded the range of such patterns to 62, most of which successfully bypass both IDEs and compilers.\footnote{More details can be found in the experienment part.\ref{unreach}}

We perform unreachable code insertion for randomly selected files in our code corpus while preserving the original function of those snippets. The inserted dead code blocks are also randomly generated. To mitigate label leakage and overfitting, we only use 32 out of the 62 patterns for training, reserving the remaining 30 patterns exclusively for testing. Additionally, we keep the original code snippets in the training corpus as hard negatives.

After combining samples of both unused and unreachable code, we construct the AIDCE dataset, which comprises Python and Java code. The AIDCE dataset supports a 3-type classification of dead code and provides auxiliary information such as the specific lines of dead code. The statistics of the AIDCE dataset are presented in \ref{tab:aidce_stats}. The train/test/dev sets are split into approximately 80\%/10\%/10\% respectively.

\begin{table}[h]
\centering
\caption{AIDCE dataset statistics}
\small
\begin{tabular}{lccc}
\toprule
 & Java & Python & \cellcolor{gray!20}Total \\
\midrule
Normal & 4427 & 8201 & \cellcolor{gray!20}12628 \\
Unused & 1853 & 2642 & \cellcolor{gray!20}4495 \\
Unreachable & 207 & 2309 & \cellcolor{gray!20}2516 \\
\cellcolor{gray!20}Total & \cellcolor{gray!20}6487 & \cellcolor{gray!20}13152 & \cellcolor{gray!20}19639 \\
\bottomrule
\end{tabular}
\label{tab:aidce_stats}
\end{table}


Since we require a model that is not just a classifier, but also a powerful model that locates, explains, and removes dead code, we turn to LLMs to generate human-readable content that facilitates dead code elimination with minimal human effort. To achieve this, we construct a more detailed and complex training set for LLM fine-tuning.
To be more specific, different from directly querying GPT-4, we provide it with accurate signals including the classification labels and location of dead code, ensuring the output quality of explanations and fixed code.

\subsubsection{Model Training} We train a pivot model to function as a quick filter and dead code locator, and an LLM as a robust and effective model for double-checking suspect code snippets. The LLM is also trained to generate detailed explanations for dead code and provide practical patches for its elimination.

\textbf{Pivot Model. }
We expect the pivot model to be slim but effective, with a high recall rate in classifying dead code. This ensures that it successfully detects nearly all dead code snippets. While false positives (normal code misclassified as dead code) may occur, the LLMs deployed afterward will perform a double-check mechanism to minimize this. In practice, we choose CodeBERT~\cite{feng2020codebert}, an encoder-only transformer-based pre-trained model, as the backbone.

\textbf{LLMs. }
Existing powerful LLMs, such as GPT-4o, can accomplish most of our tasks. However, even state-of-the-art LLMs may struggle with dead code detection due to a lack of extensive training samples specifically for dead code. Additionally, larger LLMs require more computational resources and incur higher API costs. To address these issues, we select Qwen2-7B-Instruct as our base model and utilize our curated synthetic data for supervised fine-tuning (SFT). Our synthetic data is comprised mostly of gold labels from the AIDCE dataset, ensuring accuracy, while silver labels generated by GPT-4o enhance patch generation quality. We employ the LLaMA-Factory~\cite{zheng2024llamafactory} library to train a LoRA~\cite{hulora} adapter.

\subsection{Inference Phase}

\subsubsection{Dead Code Attribution} Previous work on leveraging language models for per-line code analysis, such as static program slicing~\cite{yadavally2024learning} and fault localization~\cite{yang2024large}, typically uses line-level encoders to represent each line as an embedding vector. These models predict each line individually, ignoring the broader code context. However, dead code elimination requires a global understanding of the code, including tracking variable references and analyzing conditions across multiple lines. To address this, we designed the pivot model as a three-class classifier that processes the entire code snippet, providing a global context for dead code detection but lacking the capability to pinpoint the exact lines.

To empower the pivot model with dead code localization, we introduce a novel method, dead code attribution, which measures the effect of each line on the classification task. Adapting the concept of attribution value (or Shapley value) widely used in model interpretation~\cite{mosca2022shap, nguyen2021effectiveness}, we apply it to code analysis, marking the first use of attribution value in a code-related task to assess the contribution of each line to the model's prediction values instead of single token~\cite{li2016understanding, kim2020interpretation}.



To be specific, we train a pivot classifier $f$ receiving an $n$-line code snippet $C = \{l_1, l_2, \cdots, l_n\}$ as input, where $l_i = \{c_1^i, c_2^i, \cdots, c_{m_i}^i\}$ represents the tokens of the $i$-th line, containing a total of $m_i$ tokens. The classifier $f$ predicts among 3 possible classes: normal, unused, or unreachable, resulting in a probability vector $f(C) \in [0, 1]^3$.

Before assessing the attribution of each code line $l_i \in C$, it is essential to understand a crucial characteristic in dead code elimination: removing dead code does not affect the program's functionality, whereas removing functional code can create new dead code by eliminating references to variables or methods, causing other lines to become dead code. Thus, if we delete a specific code line $l_i$, the change in the dead code prediction can indicate whether $l_i$ contains dead code. If the prediction value for the dead code class decreases, the deleted $l_i$ was likely dead code, and vice versa. We must point out that, the removal of dead code perfectly fits the motivation for evaluating the importance of each line in classification.

We classify dead code into two types. For non-condition code lines, such as assignment and computation statements, we use the Leave-One-Out (LOO) strategy that removes each line of code:
\begin{equation}
    C_{-i} = C - \{l_i\}
\end{equation}

For condition-type code lines that can be classified into \textbf{unreachable}, like \textit{if(condition)} or \textit{while(condition)}, directly deleting them will significantly alter the code's execution logic and potentially create new dead code. Instead, we replace the conditions with a mask token from the CodeBERT model, \textit{[mask]}, making it impossible to determine the value of conditions as always true or false:
\begin{equation}
    C_{-i} = mask(C, l_i)
\end{equation}
By masking the condition, if the original condition was unreachable, masking it will prevent it from being classified as unreachable. If the condition is normal and functional, masking it will obscure its true behavior and increase the probability of misclassification as unused or unreachable. However, we only need to focus on the \textbf{decrease} of probabilities on unused and unreachable classes as mentioned above.

Thus, we define the attribution value as:
\begin{equation}
    a_{i} = max(f(C) - f(C_{-i}),\ 0)
\end{equation}

which denotes the probability difference before and after eliminating the assumed dead code line. It is clear that this value reflects the attribution of the line about being classified as dead code.

\begin{figure}[ht]
\centering
\includegraphics[width=7.35cm]{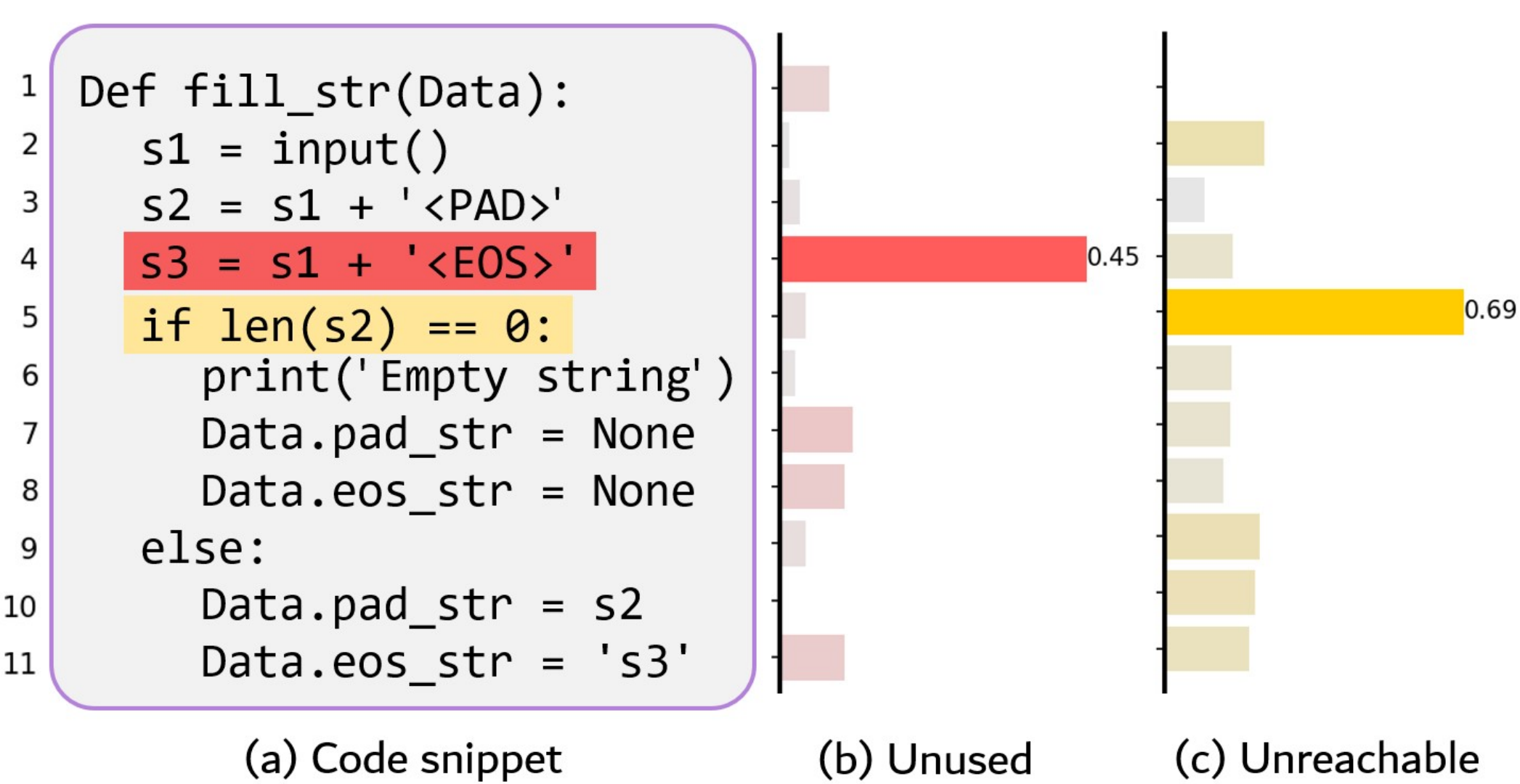}
\caption{An illustrative Python method with dead code.}
\label{fig4}
\end{figure}

We focus on the attribution value of unused and unreachable code. Hence, $a_{i}$ is regarded as a 2-dimensional vector $a_{i} \in [0, 1]^2$. The greater the value of $a_{i}$, the higher the likelihood that line $c_i$ contains dead code. This approach allows us to distribute the prediction score $f(C)$ across all lines of code in a mini-batch, enabling us to locate dead code lines without specifically training a line-level classifier. Figure \ref{fig4} presents an example of our dead code attribution algorithm.

\subsubsection{Pipeline Overview}
By enabling the precise location of dead code with our proposed dead code attribution approach, we can complete the pipeline for dead code elimination, encompassing classification, location, exploration, and patching. Specifically, our approach involves three key steps, as illustrated in Figure \ref{fig3}:

\textbf{Pivot Model Filtering.} We initiate the process with a series of tokens representing the code snippet $C = {l_1, l_2, \cdots, l_n}$. This snippet is input into our pivot model $f$, which predicts one of three categories: normal, unused, or unreachable. To reduce computational overhead for subsequent steps, we filter out samples classified as normal. During the training phase, we ensure the model achieves a high recall rate for the unused and unreachable classes, thereby ensuring that nearly all potential dead code advances to the following stages.

\textbf{Dead Code Locating.}
In this phase, we process each snippet $C$ identified as containing potential dead code. Utilizing our dead code attribution approach, we evaluate each line of code in an $n$-size mini-batch, resulting in attribution scores $\boldsymbol{a} = {a_1, a_2, \cdots, a_n}$. We focus on the probability differences for unused (the first dimension) and unreachable (the second dimension) lines, ranking them separately in descending order. For instance, when addressing unused code, we sort the first dimension of $a$ to produce a ranked list. Instead of merely returning the top-$k$ results, we employ a scaling factor $\tau$, using $\frac{\max \boldsymbol{a^1}}{\tau}$ as a soft threshold to filter out values closer to the maximum, thereby forming a smaller but more accurate candidate set of lines likely to contain dead code.

\textbf{Explaining \& Eliminating.}
Through the preceding steps, we provide large language models (LLMs) with enriched prompts regarding dead code, particularly in identifying specific dead code lines. This enables the LLMs to more accurately identify and understand dead code within code snippets, thereby enhancing the explanation and elimination process. Specifically, we input the candidate set of suspect lines as auxiliary information into a fine-tuned LLM, instructing it to output detailed information . Notably, our explanations are more comprehensive and user-friendly compared to those offered by traditional IDEs and checkers. Following this step, developers can utilize the explanation information to verify the accuracy of the detected dead code and review the generated corrections. Leveraging the detailed insights provided by the LLM, our DCE-LLM pipeline significantly reduces human effort in the dead code elimination process.

\section{Evaluation}





\subsection{Evaluation Setup}


\textbf{Dataset.} We evaluate the performance of DCE-LLM against various baselines using the test split of our proposed AIDCE dataset. Note that, for unused code, we use the reports from IDEs as the gold standard labels, assuming their accuracy to be 100\%. The test set is divided into Python and Java subsets, each further split into unused code and unreachable code categories.

\noindent \textbf{Baselines.} Considering that DCE-LLM is testing-free and light-weight which does not require any program running time, we only focus on static analyzing methods or compilers. IDEs and checkers serve as the gold standard for unused code, so we only evaluate them in the context of unreachable code detection. For Java, we select IntelliJ IDEA Java IDE and the Java Just-In-Time (JIT) compiler, which can eliminate dead code during execution. Considering Python is an interpreted programming language with dynamic semantics, we adopt PyCharm IDE and Vulture as baselines.

The second type of baseline involves using Large Language Models (LLMs).
Our evaluation includes several leading models, such as API calls to \textit{GPT-4o} and \textit{GPT-3.5-turbo}, as well as open-source LLMs like \textit{Llama3-8b-instruct} and \textit{Qwen2-7B-instruct}. Code-specific LLMs such as \textit{Deepseek-coder-6.7B} and \textit{Mamba-Codestral-7B} are also evaluated.
Due to our limited computing resources, we only assess LLMs up to a maximum size of 8B parameters. For GPT-4o, we also evaluate it with few-shot prompt.

\label{metric}
\noindent \textbf{Metrics.} We first assess the classification performance of different methods with standard multi-task classification metrics, including accuracy, precision, recall, and F1-Score for the unused and unreachable classes.
Moreover, we assess the quality of generated content with human evaluation, including explanations and repaired code due to the absence of gold standard of generated contents.


\begin{table*}[h!]
\centering
\renewcommand{\arraystretch}{1} 
\setlength{\tabcolsep}{5.5pt} 
\small
\begin{tabular}{c ccc ccc ccc c}
\toprule
\multirow{2}{*}{\textbf{Approach}} & \multicolumn{3}{c}{\textbf{Unused}} & \multicolumn{3}{c}{\textbf{Unreachable}} & \multicolumn{3}{c}{\textbf{Normal}} & \multirow{2}{*}{\textbf{Accuracy}} \\
\cmidrule(lr){2-10}
 & \textbf{R} & \textbf{P} & \textbf{F1} & \textbf{R} & \textbf{P} & \textbf{F1} & \textbf{R} & \textbf{P} & \textbf{F1} & \\
\midrule

\multirow{1}{*}{IDEs} & 100.0 & 100.0 & 100.0 & 5.98 & 100.0 & 9.48 & 100.0 & 77.36 & 87.24 & 87.89\\

\midrule
\midrule
\multirow{1}{*}{Llama-3-8B} & 39.78 & 20.97 & 27.46 & 44.40 & 29.89 & 35.73 & 40.39 & 75.33 & 52.58 & 38.34 \\

\midrule
\multirow{1}{*}{Qwen2-8B} & 8.31 & 41.57 & 13.86 & 68.46 & 80.88 & 74.16 & 98.29 & 75.00 & 85.08 & 71.55\\

\midrule
\multirow{1}{*}{GPT-3.5} & 7.64 & 38.20 & 12.73 & 59.75 & 70.59 & 64.72 & 94.24 & 74.80 & 83.40 & 69.59\\

\midrule
\multirow{1}{*}{GPT-4o} & 47.19 & 36.21 & 40.98 & 82.99 & 52.08 & 64.00 & 70.97 & 84.06 & 76.96 & 60.97\\

\midrule
\multirow{1}{*}{GPT-4o (3-shot)} & 28.31 & 45.48 & 34.90 & 36.51 & 58.66 & 45.01 & 86.69 & 76.30 & 81.16 & 67.79\\

\midrule
\midrule
\multirow{1}{*}{Codellama-7B} & 96.40 & 23.87 & 38.26 & 45.22 & 17.44 & 25.17 & 2.33 & 75.00 & 4.52 & 16.07 \\

\midrule
\multirow{1}{*}{Deepseek-coder-6.7B} & 0.0 & 0.0 & 0.0 & 2.85 & 9.09 & 4.34 & 95.71 & 70.27 & 81.04 & 67.67 \\

\midrule
\multirow{1}{*}{Mamba-Codestral-7B} & 7.19 & 40.00 & 12.19 & 6.22 & 83.33 & 11.58 & 98.91 & 70.29 & 82.18 & 67.53 \\

\midrule
\multirow{1}{*}{Qwen-2.5-coder-32B} & 13.71 & 36.75 & 19.97 & 80.50 & 74.90 & 77.60 & 93.70 & 77.28 & 84.97 & 70.44\\

\midrule
\midrule
\rowcolor{gray!20} \multirow{1}{*}{DCE-LLM (Ours)} & \cellcolor{gray!20}\textbf{93.71} & \cellcolor{gray!20}\textbf{94.34} & \cellcolor{gray!20}\textbf{94.02} & \cellcolor{gray!20}\textbf{95.85} & \cellcolor{gray!20}\textbf{97.47} & \cellcolor{gray!20}\textbf{96.65} & \cellcolor{gray!20}\textbf{99.69} & \cellcolor{gray!20}\textbf{99.07} & \cellcolor{gray!20}\textbf{99.38} & \cellcolor{gray!20}\textbf{96.40} \\

\bottomrule
\end{tabular}
\caption{Comparison of various baselines}
\label{tab3}
\end{table*}

\subsection{Classification Performance}
\label{unreach}

Despite their effectiveness in detecting unused code, IDEs, checkers, and compilers perform poorly when faced with unreachable adversarial attacks. In addition to the patterns designed in DaK~\cite{gao2023discrete}, we have expanded the number of patterns to 62. Table \ref{taba} presents several of these patterns. We use \textit{PyCharm} and \textit{IntelliJ IDEA} as IDEs for checking unreachable code, denoted by brown {\textcolor{brown}{\ding{51}}}/{\textcolor{brown}{\ding{55}}} symbols for Python and Java, respectively. Additionally, we employ the \textit{Vulture} checker (for Python) and the \textit{JIT} compiler (for Java) to evaluate the same pattern set, with their results indicated by blue {\textcolor{darkblue}{\ding{51}}}/{\textcolor{darkblue}{\ding{55}}} symbols. Remarkably, even IntelliJ IDEA, one of the most advanced and widely used IDEs, successfully identifies only 18 out of the 62 patterns, which is less than 30\%. Other static analysis tools such as UCDetector and J2ObjC show weaker detection capabilities compared to IntelliJ IDEA. The UCDetector detected 22 out of 100 unused code cases but failed to detect any of the 62 unreachable patterns.
Considering that those tools are based on rules, we can assume that the precision value $P$ for unreachable code detection is $1$, and the recall rate equals the checked ratio as we randomly select inserted patterns.


For learning-based approaches, Table \ref{tab3} details the experimental results comparing DCE-LLM with other LLMs. IDEs for different languages are ensembled into a powerful baseline. Four foundation LLMs and 3 code-specific LLMs are evaluated alongside our method. Except for the \textit{unused} split where IDEs serve as gold labels, we observe that DCE-LLM achieves the highest scores across all metrics in Python and Java code corpus
\footnote{Detailed performance in Python/Java split are provided in Appendix.}
, including foundation LLMs and code-specific LLMs.

Due to the unbalanced label distribution, a model can easily achieve high accuracy by predicting all code as normal. Even GPT-4o exhibits relatively lower accuracy compared to GPT-3.5. Thus, focusing on the detection of dead code, the recall and F1 scores for unused and unreachable samples are more critical metrics. All LLMs struggle to detect unused code, with recall rates below 50\%. The global characteristics of unused code detection challenge the long-form memorization ability of LLMs, which remains an active area of research. DCE-LLM, however, achieves a recall rate of \textit{93.71\%}, marking a substantial improvement over other baselines.

For unreachable code detection, LLMs demonstrate an advantage in analyzing and computing representations in conditional statements. While IDEs rarely detect unreachable conditions, GPT-4o identifies unreachable code with a recall rate of 82.99\%, and Qwen2-8B achieves a precision of 80.88\%.
Our proposed DCE-LLM surpasses these models, with a leading recall rate of 95.85\% and a precision rate of 97.47\%.

From our study, the most challenging pattern for Python involves type-based unreachable conditions (e.g., \texttt{isInstance(int(a))}). Additionally, the model occasionally misses several dead code lines when multiple dead code lines appear together.


\begin{figure}[ht]
\centering
\includegraphics[width=7.5cm]{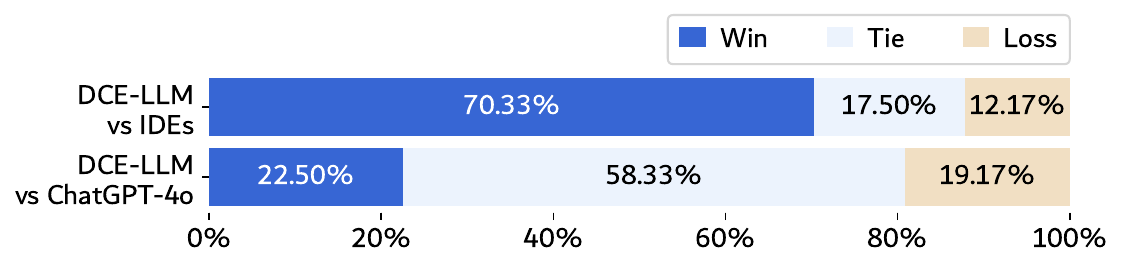}
\caption{Human evaluation on explanation generation.}
\label{fig5}
\end{figure}

\begin{figure}[htbp]
\centering
\includegraphics[width=7.5cm]{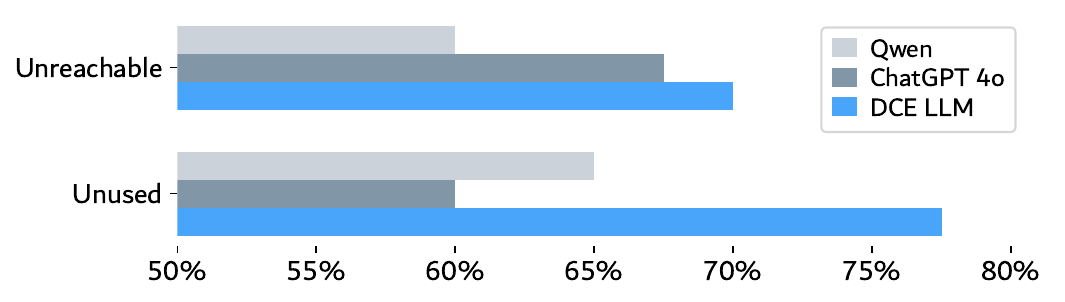}
\caption{Human evaluation on code repairing.}
\label{fig6}
\end{figure}

\subsection{Generation Quality}

To measure the quality of generated contents from DCE-LLM, we conduct a human evaluation of both explanations and repaired code. Three experienced programmers serve as voluntary annotators on 20 unused codes and 20 unreachable codes.


For explanation, annotators rate each output as worse, similar, or better than the baseline. As depicted in Figure \ref{fig5}, our method outperformed the baseline methods in explanation.

For code repairing, we focus on two major points: (1) if the code maintains the same function and (2) if all dead code is removed. Only samples satisfying both two conditions are positive. As shown in Figure \ref{fig6}, DCE-LLM surpasses both Qwen and GPT-4o in the evaluation.

Our study revealed that IDEs provide only template-based explanations for unused code, whereas our model generates more diverse explanations. Furthermore, due to the high-quality synthetic data and attribution techniques used, our model can accurately locate and explain dead code lines, leading to more applicable code patches.



\subsection{Ablation Study}
To understand the impact of the major components in DCE-LLM, we separate our pivot model and generation model to test them as individual classifiers. Additionally, we explore minor modifications of the pipeline. Our ablation results are presented in Table \ref{tab4}, demonstrating the effectiveness of the essentials in our pipeline.

\begin{table}[h]
\centering
\small
\renewcommand{\arraystretch}{1.2}
\resizebox{\columnwidth}{!}{
\begin{tabular}{l ccc ccc}

\toprule
\multirow{2}{*}{\textbf{Approach}} & \multicolumn{3}{c}{\textbf{Unused}} & \multicolumn{3}{c}{\textbf{Unreachable}} \\
\cmidrule(lr){2-7}
 & \textbf{R} & \textbf{P} & \textbf{F1} & \textbf{R} & \textbf{P} & \textbf{F1} \\

\midrule

\multirow{1}{*}{\raggedright \cellcolor{gray!20}\ \ DCE-LLM} & \cellcolor{gray!20}93.71 & \cellcolor{gray!20}94.34 & \cellcolor{gray!20}94.02 & \cellcolor{gray!20}95.85 & \cellcolor{gray!20}97.47 & \cellcolor{gray!20}96.65 \\

\multirow{1}{*}{\raggedright - Pivot Model} & 51.46 & 87.07 & 64.69 & 84.65 & 99.02 & 91.28 \\
\multirow{1}{*}{\raggedright - LLM} & 94.61 & 92.32 & 93.45 & 94.19 & 99.13 & 96.60 \\
\multirow{1}{*}{\raggedright - SFT} & 48.52 & 96.57 & 64.59 & 98.58 & 50.73 & 66.99 \\
\multirow{1}{*}{\raggedright - Attribution} & 83.37 & 98.40 & 90.27 & 91.30 & 95.85 & 93.52 \\
\bottomrule
\end{tabular}}
\caption{Ablation study results of DCE-LLM}
\label{tab4}
\end{table}


\textit{Without Pivot Model, }we directly prompt our fine-tuned LLM to perform dead code elimination. This results in a significant decline in performance, with a drop of about 30\% in the F1 score for unused code detection and a slight decrease for unreachable code. The pivot model plays a crucial role in retrieving most dead code snippets from the corpus.

\textit{Without LLM, }the classification results show only minor changes. However, the ability to generate explanations and repair code is completely lost.

\textit{Without SFT, }we replace our fine-tuned LLM with Qwen2-7B-instruct model. This substitution results in a 30\% drop in the F1 score for both splits. Without the specialized training on our well-annotated corpus, the LLM cannot neither maintaining high classification performance or generate high-quality contents.

\textit{Without Dead Code Attribution, }the LLM loses precise guidance on where dead code is located, leading to confusion in the dead code elimination process. In this configuration, all metrics decrease by approximately 3\%.


\subsection{Programming Language Generalization}
To test the cross-language generalization of DCE-LLM, we annotated 275 Golang files as a test set. We applied a unreachable pattern attack on this set, resulting in 100 unused and 98 unreachable samples (some of which overlap).
The Golang dataset is simpler than Java/Python with less random generated variables and obfuscations.
We then leveraged DCE-LLM for dead code detection on this corpus, and the results are reported in Table \ref{tab5}. From the table, we can observe that even when facing an unseen programming language, DCE-LLM still showcases strong performance in detecting unused code.
For unreachable code, the performance remains robust, although slightly lower compared to unused code detection. This indicates that while DCE-LLM effectively generalizes to new programming languages, there is still room for improvement in handling more complex unreachable code scenarios. Performance on both unused and unreachable code outperforms GPT-4o.

\begin{table}[h]
\centering
\small
\renewcommand{\arraystretch}{1.2}
\resizebox{\columnwidth}{!}{
\begin{tabular}{l ccc ccc}

\toprule
\multirow{2}{*}{\textbf{Approach}} & \multicolumn{3}{c}{\textbf{Unused}} & \multicolumn{3}{c}{\textbf{Unreachable}} \\
\cmidrule(lr){2-7}
 & \textbf{R} & \textbf{P} & \textbf{F1} & \textbf{R} & \textbf{P} & \textbf{F1} \\

\midrule

\multirow{1}{*}{GPT-4o} & 73.00 & 62.93 & 67.59 & 80.61 & 83.16 & 81.87 \\
\multirow{1}{*}{\cellcolor{gray!20}DCE-LLM} & \cellcolor{gray!20}85.00 & \cellcolor{gray!20}96.59 & \cellcolor{gray!20}90.42 & \cellcolor{gray!20}90.00 & \cellcolor{gray!20}91.84 & \cellcolor{gray!20}90.91 \\

\bottomrule
\end{tabular}}
\caption{Performance of DCE-LLM on Golang}
\label{tab5}
\end{table}




\section{Conclusion}

In this paper, we proposed DCE-LLM, the first learning-based solution for the dead code elimination task that leverages a pivot model for filtering and recall, and an LLM for explanation and code fixing. The LLM is empowered with location information provided by the dead code attribution technique.
By fine-tuning both the pivot model and LLM on the first large-scale dead code dataset, we successfully constructed a complete pipeline for dead code elimination, encompassing classification, location, explanation, and repair. Our proposed framework demonstrates superior performance in both classification and generation tasks. Additionally, the versatility of DCE-LLM enables generalization on other programming languages.







\section*{Limitation}
Firstly, the pivot model introduces a relatively limited input length. The input length is 512 for CodeBERT, while programs can easily exceed. Thus, this work focus on line-level dead code elimination instead of method-level. We plan to support longer contexts and complex control flows in the future.

Secondly, the study of prompting techniques can be insufficient. We only compares the standard CoT prompt with ours. We did not studied prompt engineering and devising more sophisticated in-context examples which is beyond the scope of this work. We also believe that no prompting method can bridge the over 30\% F1 performance gap.

Lastly, as a learning-based method, DCE-LLM is not a complete that detects all dead code. We plan to conduct case study on failure cases to improve our model in the future.

\bibliography{custom}

\begin{thebibliography}{32}
\providecommand{\natexlab}[1]{#1}

\bibitem[{AlAbwaini et~al.(2018)AlAbwaini, Aldaaje, Jaber, Abdallah, and
  Tamimi}]{alabwaini2018using}
Nour AlAbwaini, Amal Aldaaje, Tamara Jaber, Mohammad Abdallah, and Abdelfatah
  Tamimi. 2018.
\newblock Using program slicing to detect the dead code.
\newblock In \emph{2018 8th International Conference on Computer Science and
  Information Technology (CSIT)}, pages 230--233. IEEE.

\bibitem[{Bhatia et~al.(2024)Bhatia, Qiu, Hasabnis, Seshia, and
  Cheung}]{bhatia2024verified}
Sahil Bhatia, Jie Qiu, Niranjan Hasabnis, Sanjit~A. Seshia, and Alvin Cheung.
  2024.
\newblock \href {https://openreview.net/forum?id=spwE9sLrfg} {Verified code
  transpilation with {LLM}s}.
\newblock In \emph{The Thirty-eighth Annual Conference on Neural Information
  Processing Systems}.

\bibitem[{Boomsma et~al.(2012)Boomsma, Hostnet, and Gross}]{boomsma2012dead}
Hidde Boomsma, BV~Hostnet, and Hans-Gerhard Gross. 2012.
\newblock Dead code elimination for web systems written in php: Lessons learned
  from an industry case.
\newblock In \emph{2012 28th IEEE International Conference on Software
  Maintenance (ICSM)}, pages 511--515. IEEE.

\bibitem[{Chen et~al.(1998)Chen, Gansner, and Koutsofios}]{chen1998c++}
Yih-Fam Chen, Emden~R Gansner, and Eleftherios Koutsofios. 1998.
\newblock A c++ data model supporting reachability analysis and dead code
  detection.
\newblock \emph{IEEE Transactions on Software Engineering}, 24(9):682--694.

\bibitem[{Feng et~al.(2020)Feng, Guo, Tang, Duan, Feng, Gong, Shou, Qin, Liu,
  Jiang et~al.}]{feng2020codebert}
Zhangyin Feng, Daya Guo, Duyu Tang, Nan Duan, Xiaocheng Feng, Ming Gong, Linjun
  Shou, Bing Qin, Ting Liu, Daxin Jiang, et~al. 2020.
\newblock Codebert: A pre-trained model for programming and natural languages.
\newblock In \emph{Findings of the Association for Computational Linguistics:
  EMNLP 2020}, pages 1536--1547.

\bibitem[{Gao et~al.(2023)Gao, Wang, and Wang}]{gao2023discrete}
Fengjuan Gao, Yu~Wang, and Ke~Wang. 2023.
\newblock Discrete adversarial attack to models of code.
\newblock \emph{Proceedings of the ACM on Programming Languages},
  7(PLDI):172--195.

\bibitem[{Han et~al.(2024)Han, Kim, Yoo, Lee, and Hwang}]{han2024archcode}
Hojae Han, Jaejin Kim, Jaeseok Yoo, Youngwon Lee, and Seung-won Hwang. 2024.
\newblock Archcode: Incorporating software requirements in code generation with
  large language models.
\newblock In \emph{Proceedings of the 62nd Annual Meeting of the Association
  for Computational Linguistics (Volume 1: Long Papers)}, pages 13520--13552.

\bibitem[{Hu et~al.(2022)Hu, yelong shen, Wallis, Allen-Zhu, Li, Wang, Wang,
  and Chen}]{hulora}
Edward~J Hu, yelong shen, Phillip Wallis, Zeyuan Allen-Zhu, Yuanzhi Li, Shean
  Wang, Lu~Wang, and Weizhu Chen. 2022.
\newblock \href {https://openreview.net/forum?id=nZeVKeeFYf9} {Lo{RA}: Low-rank
  adaptation of large language models}.
\newblock In \emph{International Conference on Learning Representations}.

\bibitem[{IntelliJ(2011)}]{intellij2011most}
IDEA IntelliJ. 2011.
\newblock the most intelligent java ide.
\newblock \emph{JetBrains [online].[cit. 2016-02-23]. Dostupn{\'e} z:
  https://www. jetbrains. com/idea/\# chooseYourEdition}.

\bibitem[{Jain et~al.()Jain, Zhang, Chiang, Gonzalez, Sen, and
  Stoica}]{jainllm}
Naman Jain, Tianjun Zhang, Wei-Lin Chiang, Joseph~E Gonzalez, Koushik Sen, and
  Ion Stoica.
\newblock Llm-assisted code cleaning for training accurate code generators.
\newblock In \emph{The Twelfth International Conference on Learning
  Representations}.

\bibitem[{Khare et~al.(2023)Khare, Dutta, Li, Solko-Breslin, Alur, and
  Naik}]{khare2023understanding}
Avishree Khare, Saikat Dutta, Ziyang Li, Alaia Solko-Breslin, Rajeev Alur, and
  Mayur Naik. 2023.
\newblock Understanding the effectiveness of large language models in detecting
  security vulnerabilities.
\newblock \emph{arXiv preprint arXiv:2311.16169}.

\bibitem[{Kim et~al.(2020)Kim, Yi, Kim, and Yoon}]{kim2020interpretation}
Siwon Kim, Jihun Yi, Eunji Kim, and Sungroh Yoon. 2020.
\newblock Interpretation of nlp models through input marginalization.
\newblock In \emph{Proceedings of the 2020 Conference on Empirical Methods in
  Natural Language Processing (EMNLP)}, pages 3154--3167.

\bibitem[{Kwon et~al.(2023)Kwon, Li, Zhuang, Sheng, Zheng, Yu, Gonzalez, Zhang,
  and Stoica}]{kwon2023efficient}
Woosuk Kwon, Zhuohan Li, Siyuan Zhuang, Ying Sheng, Lianmin Zheng, Cody~Hao Yu,
  Joseph Gonzalez, Hao Zhang, and Ion Stoica. 2023.
\newblock Efficient memory management for large language model serving with
  pagedattention.
\newblock In \emph{Proceedings of the 29th Symposium on Operating Systems
  Principles}, pages 611--626.

\bibitem[{Li et~al.(2024)Li, Zhou, and Shen}]{li2024rewriting}
Haochen Li, Xin Zhou, and Zhiqi Shen. 2024.
\newblock Rewriting the code: A simple method for large language model
  augmented code search.
\newblock \emph{arXiv preprint arXiv:2401.04514}.

\bibitem[{Li et~al.(2016)Li, Monroe, and Jurafsky}]{li2016understanding}
Jiwei Li, Will Monroe, and Dan Jurafsky. 2016.
\newblock Understanding neural networks through representation erasure.
\newblock \emph{arXiv preprint arXiv:1612.08220}.

\bibitem[{Malavolta et~al.(2023)Malavolta, Nirghin, Scoccia, Romano, Lombardi,
  Scanniello, and Lago}]{malavolta2023javascript}
Ivano Malavolta, Kishan Nirghin, Gian~Luca Scoccia, Simone Romano, Salvatore
  Lombardi, Giuseppe Scanniello, and Patricia Lago. 2023.
\newblock Javascript dead code identification, elimination, and empirical
  assessment.
\newblock \emph{IEEE Transactions on Software Engineering}, 49(7):3692--3714.

\bibitem[{Mosca et~al.(2022)Mosca, Szigeti, Tragianni, Gallagher, and
  Groh}]{mosca2022shap}
Edoardo Mosca, Ferenc Szigeti, Stella Tragianni, Daniel Gallagher, and Georg
  Groh. 2022.
\newblock Shap-based explanation methods: a review for nlp interpretability.
\newblock In \emph{Proceedings of the 29th international conference on
  computational linguistics}, pages 4593--4603.

\bibitem[{Nguyen et~al.(2021)Nguyen, Kim, and Nguyen}]{nguyen2021effectiveness}
Giang Nguyen, Daeyoung Kim, and Anh Nguyen. 2021.
\newblock The effectiveness of feature attribution methods and its correlation
  with automatic evaluation scores.
\newblock \emph{Advances in Neural Information Processing Systems},
  34:26422--26436.

\bibitem[{Puri et~al.(2021)Puri, Kung, Janssen, Zhang, Domeniconi, Zolotov,
  Dolby, Chen, Choudhury, Decker, Thost, Buratti, Pujar, Ramji, Finkler,
  Malaika, and Reiss}]{puri2021codenet}
Ruchir Puri, David~S Kung, Geert Janssen, Wei Zhang, Giacomo Domeniconi,
  Vladimir Zolotov, Julian Dolby, Jie Chen, Mihir Choudhury, Lindsey Decker,
  Veronika Thost, Luca Buratti, Saurabh Pujar, Shyam Ramji, Ulrich Finkler,
  Susan Malaika, and Frederick Reiss. 2021.
\newblock \href {https://openreview.net/forum?id=6vZVBkCDrHT} {Codenet: A
  large-scale {AI} for code dataset for learning a diversity of coding tasks}.
\newblock In \emph{Thirty-fifth Conference on Neural Information Processing
  Systems Datasets and Benchmarks Track (Round 2)}.

\bibitem[{Romano and Scanniello(2018)}]{romano2018exploring}
Simone Romano and Giuseppe Scanniello. 2018.
\newblock Exploring the use of rapid type analysis for detecting the dead
  method smell in java code.
\newblock In \emph{2018 44th Euromicro conference on software engineering and
  advanced applications (SEAA)}, pages 167--174. IEEE.

\bibitem[{Romano et~al.(2016)Romano, Scanniello, Sartiani, and
  Risi}]{romano2016graph}
Simone Romano, Giuseppe Scanniello, Carlo Sartiani, and Michele Risi. 2016.
\newblock A graph-based approach to detect unreachable methods in java
  software.
\newblock In \emph{Proceedings of the 31st Annual ACM symposium on applied
  computing}, pages 1538--1541.

\bibitem[{Romano et~al.(2020)Romano, Vendome, Scanniello, and
  Poshyvanyk}]{8370748}
Simone Romano, Christopher Vendome, Giuseppe Scanniello, and Denys Poshyvanyk.
  2020.
\newblock \href {https://doi.org/10.1109/TSE.2018.2842781} {A multi-study
  investigation into dead code}.
\newblock \emph{IEEE Transactions on Software Engineering}, 46(1):71--99.

\bibitem[{Seipp()}]{Vulture}
Jendrik Seipp.
\newblock Vulture, find dead python code.
\newblock \url{https://github.com/jendrikseipp/vulture/}.

\bibitem[{Tang et~al.(2024{\natexlab{a}})Tang, Hu, Zhou, Zhong, Zheng, Si, and
  Ellis}]{tang2024code}
Hao Tang, Keya Hu, Jin~Peng Zhou, Si~Cheng Zhong, Wei-Long Zheng, Xujie Si, and
  Kevin Ellis. 2024{\natexlab{a}}.
\newblock \href {https://openreview.net/forum?id=o863gX6DxA} {Code repair with
  {LLM}s gives an exploration-exploitation tradeoff}.
\newblock In \emph{The Thirty-eighth Annual Conference on Neural Information
  Processing Systems}.

\bibitem[{Tang et~al.(2024{\natexlab{b}})Tang, Kim, Song, Lothritz, Li, Ezzini,
  Tian, Klein, and Bissyand{\'e}}]{Tang2024CodeAgentAC}
Xunzhu Tang, Kisub Kim, Yewei Song, Cedric Lothritz, Bei Li, Saad Ezzini, Haoye
  Tian, Jacques Klein, and T{\'e}gawend{\'e}~F. Bissyand{\'e}.
  2024{\natexlab{b}}.
\newblock Codeagent: Autonomous communicative agents for code review.

\bibitem[{Wang et~al.(2024)Wang, Zhang, Su, Xu, and
  Zhang}]{wang-etal-2024-sanitizing}
Chengpeng Wang, Wuqi Zhang, Zian Su, Xiangzhe Xu, and Xiangyu Zhang. 2024.
\newblock \href {https://doi.org/10.18653/v1/2024.findings-emnlp.217}
  {Sanitizing large language models in bug detection with data-flow}.
\newblock In \emph{Findings of the Association for Computational Linguistics:
  EMNLP 2024}, pages 3790--3805, Miami, Florida, USA. Association for
  Computational Linguistics.

\bibitem[{Wang et~al.(2017)Wang, Zhang, Zhao, and Chen}]{wang2017dead}
Xing Wang, Yingzhou Zhang, Lian Zhao, and Xinghao Chen. 2017.
\newblock Dead code detection method based on program slicing.
\newblock In \emph{2017 International Conference on Cyber-Enabled Distributed
  Computing and Knowledge Discovery (CyberC)}, pages 155--158. IEEE.

\bibitem[{Wei et~al.()Wei, Durrett, and Dillig}]{weicoeditor}
Jiayi Wei, Greg Durrett, and Isil Dillig.
\newblock Coeditor: Leveraging repo-level diffs for code auto-editing.
\newblock In \emph{The Twelfth International Conference on Learning
  Representations}.

\bibitem[{Yadavally et~al.(2024)Yadavally, Li, Wang, and
  Nguyen}]{yadavally2024learning}
Aashish Yadavally, Yi~Li, Shaohua Wang, and Tien~N Nguyen. 2024.
\newblock A learning-based approach to static program slicing.
\newblock \emph{Proceedings of the ACM on Programming Languages},
  8(OOPSLA1):83--109.

\bibitem[{Yang et~al.(2024)Yang, Le~Goues, Martins, and
  Hellendoorn}]{yang2024large}
Aidan~ZH Yang, Claire Le~Goues, Ruben Martins, and Vincent Hellendoorn. 2024.
\newblock Large language models for test-free fault localization.
\newblock In \emph{Proceedings of the 46th IEEE/ACM International Conference on
  Software Engineering}, pages 1--12.

\bibitem[{Zhao et~al.(2024)Zhao, Huang, Ma, Li, Zhang, Jiang, Liu, Zhu, and
  Su}]{zhao-etal-2024-repair}
Yuze Zhao, Zhenya Huang, Yixiao Ma, Rui Li, Kai Zhang, Hao Jiang, Qi~Liu, Linbo
  Zhu, and Yu~Su. 2024.
\newblock \href {https://doi.org/10.18653/v1/2024.findings-acl.973}
  {{R}e{P}air: Automated program repair with process-based feedback}.
\newblock In \emph{Findings of the Association for Computational Linguistics:
  ACL 2024}, pages 16415--16429, Bangkok, Thailand. Association for
  Computational Linguistics.

\bibitem[{Zheng et~al.(2024)Zheng, Zhang, Zhang, Ye, Luo, Feng, and
  Ma}]{zheng2024llamafactory}
Yaowei Zheng, Richong Zhang, Junhao Zhang, Yanhan Ye, Zheyan Luo, Zhangchi
  Feng, and Yongqiang Ma. 2024.
\newblock \href {http://arxiv.org/abs/2403.13372} {Llamafactory: Unified
  efficient fine-tuning of 100+ language models}.
\newblock In \emph{Proceedings of the 62nd Annual Meeting of the Association
  for Computational Linguistics (Volume 3: System Demonstrations)}, Bangkok,
  Thailand. Association for Computational Linguistics.

\end{thebibliography}

\newpage
\appendix

\section{Dataset Details}
The AIDCE dataset is divided into training, validation, and test subsets with 15,043, 1,889, and 1,891 samples, respectively. To accurately represent the dead code rate from the original CodeNet dataset while including unreachable codes, we set the ratio of normal to unused to unreachable code at around 4:1:1. Some overlapped samples (around 5\%) that contain both unused and unreachable code also exist.

\begin{table}[ht]
    \centering
    \begin{tabular}{>{\centering\arraybackslash}m{0.8cm} >{\centering\arraybackslash}m{2cm} >{\centering\arraybackslash}m{2.8cm}}
        \toprule
        \multicolumn{2}{c}{\textbf{Fields}} & \textbf{Source} \\
        \midrule
        \multirow{2}{*}{Input} & Source code & Dataset \\
                               & Suspect lines & Pivot Model\\
        \cmidrule(lr){1-3}
        \multirow{5}{*}{Output} & Dead code & Dataset \\
                                & Line number & Dataset \\
                                & Type & Dataset \\
                                & Explanation & Dataset \& GPT-4o \\
                                & Fixed code & Source Code \& GPT-4o \\
        \bottomrule
    \end{tabular}
    \caption{Source of fields in SFT data for training LLMs}
    \label{taba1}
\end{table}

For SFT dataset, We collect 1,500 samples with a ratio of normal to unused to unreachable code at about 1:1:1 as our pivot model can filter out most normal code snippets before passing them into LLMs. Table \ref{taba1} provides details of source of the fields for SFT data.

\section{Model Training}
 For CodeBERT model, we a classification head on top of CodeBERT. The model is fine-tuned to predict the class of code snippets among three categories. We set the batch size to 16 and the learning rate to 5e-5. The model is fine-tuned using the Adam optimizer for up to 3 epochs, with an early stopping mechanism to prevent overfitting.

 For LLM training, we employ the LLaMA-Factory~\cite{zheng2024llamafactory} WEB GUI, we fine-tuned a LoRA~\cite{hulora} for the Qwen2-7B model with the following settings: learning rate of 5e-5, 10 epochs, a cosine learning rate scheduler, LoRA rank of 8, and alpha of 16. Additionally, we used DeepSpeed ZeRO-2 for efficient GPU memory management.

\section{Experimental Environment}
The entire training and inference phases are conducted using an AMD Ryzen 9 5950X CPU running Ubuntu 23.04 with 128GB RAM and 2 NVIDIA RTX 4090 24GB GPUs. VLLM framework~\cite{kwon2023efficient} is applied for serving our local LLMs. For relatively stable and consistent generation results, we set $temperature = 0.1$. The $max\_token$ value is aligned with the default value of backbone LLMs. In our experiments, adjusting $\tau$ allows us to extract a small number of code lines while maintaining a high recall rate for dead code location. We set the soft threshold parameter $\tau = 2$.

\section{Detailed Performance}
We present the detailed performance in Tab. \ref{taba3}.

\begin{table*}[h!]
\centering
\renewcommand{\arraystretch}{1.15} 
\setlength{\tabcolsep}{3.5pt} 
\small
\caption{Comparison of various baselines (detailed)}
\begin{tabular}{c  c ccc ccc ccc c}
\toprule
\multirow{2}{*}{\textbf{Approach}} & \multirow{2}{*}{\textbf{Language}}  & \multicolumn{3}{c}{\textbf{Unused}} & \multicolumn{3}{c}{\textbf{Unreachable}} & \multicolumn{3}{c}{\textbf{Normal}} & \multirow{2}{*}{\textbf{Accuracy}} \\
\cmidrule(lr){3-11}

 &  & \textbf{R} & \textbf{P} & \textbf{F1} & \textbf{R} & \textbf{P} & \textbf{F1} & \textbf{R} & \textbf{P} & \textbf{F1} & \\
\midrule

\multirow{3}{*}{IDEs} & Python  & 100.0 & 100.0 & 100.0 & 0 & 100.0 & 0  & 100.0 & 81.40 & 89.73 & 84.28\\
 & Java  & 100.0 & 100.0 & 100.0 & 23.08 & 100.0 & 35.49  & 100.0 & 91.98 & 95.82 & 94.19\\
 & \cellcolor{gray!20}Overall  & \cellcolor{gray!20}100.0 & \cellcolor{gray!20}100.0 & \cellcolor{gray!20}100.0 & \cellcolor{gray!20}5.98 & \cellcolor{gray!20}100.0 & \cellcolor{gray!20}9.48 &  \cellcolor{gray!20}100.0 & \cellcolor{gray!20}77.36 & \cellcolor{gray!20}87.24 & \cellcolor{gray!20}87.89\\

\midrule
\midrule

\multirow{3}{*}{Llama-3-8B} &Python   & 43.06 & 15.02 & 22.28 & 34.39 & 29.02 & 31.48 & 31.96 &  73.54 & 44.56 & 32.70\\
 & Java   & 36.68 & 37.33 & 37.00 & 80.77 & 31.34 & 45.16 & 55.56 & 77.27 & 64.64 & 48.19\\
 & \cellcolor{gray!20}Overall & \cellcolor{gray!20}39.78 & \cellcolor{gray!20}20.97 & \cellcolor{gray!20}27.46 & \cellcolor{gray!20}44.40 & \cellcolor{gray!20}29.89 & \cellcolor{gray!20}35.73 &\cellcolor{gray!20}40.39 & \cellcolor{gray!20}75.33 & \cellcolor{gray!20}52.58  & \cellcolor{gray!20}38.34 \\

\midrule

\multirow{3}{*}{Qwen2-8B} & Python  & 10.65 & 32.39 & 16.03 & 68.25 & 83.23 & 75.00 & 98.18 & 77.68 & 86.74 & 73.88\\
 & Java   & 6.11 & 77.78 & 11.34 & 69.23 & 73.47 & 71.29 & 98.47 & 70.63 & 82.26 & 67.49\\
 & \cellcolor{gray!20}Overall   & \cellcolor{gray!20}8.31 & \cellcolor{gray!20}41.57 & \cellcolor{gray!20}13.86 & \cellcolor{gray!20}68.46 & \cellcolor{gray!20}80.88 & \cellcolor{gray!20}74.16  & \cellcolor{gray!20}98.29 & \cellcolor{gray!20}75.00 & \cellcolor{gray!20}85.08 & \cellcolor{gray!20}71.55\\
\midrule

\multirow{3}{*}{GPT-3.5}  & Python   & 12.04 & 36.62 & 18.12 & 56.08 & 70.20 & 62.35  & 93.46 & 77.35 & 84.65 & 72.55\\
 & Java   & 3.49 & 44.44 & 6.48 & 73.08 & 71.70 & 72.38  & 95.64 & 70.69 & 81.30 & 64.44\\
 & \cellcolor{gray!20}Overall  & \cellcolor{gray!20}7.64 & \cellcolor{gray!20}38.20 & \cellcolor{gray!20}12.73 & \cellcolor{gray!20}59.75 & \cellcolor{gray!20}70.59 & \cellcolor{gray!20}64.72  & \cellcolor{gray!20}94.24 & \cellcolor{gray!20}74.80 & \cellcolor{gray!20}83.40 & \cellcolor{gray!20}69.59\\

\midrule

\multirow{3}{*}{GPT-4o} & Python & 68.06 & 33.26 & 44.68 & 84.66 & 57.97 & 68.82 & 68.77 & 91.47 & 78.51 & 62.90 \\
 & Java  & 27.51 & 45.65 & 34.33 & 76.92 & 37.04 & 50.00   & 74.95 & 74.14 & 74.54 & 57.62\\
 & \cellcolor{gray!20}Overall  & \cellcolor{gray!20}47.19 & \cellcolor{gray!20}36.21 & \cellcolor{gray!20}40.98 & \cellcolor{gray!20}82.99 & \cellcolor{gray!20}52.08 & \cellcolor{gray!20}64.00   & \cellcolor{gray!20}70.97 & \cellcolor{gray!20}84.06 & \cellcolor{gray!20}76.96 & \cellcolor{gray!20}60.97\\

\midrule

\multirow{3}{*}{GPT-4o (3-shot)} & Python & 41.20 & 44.05 & 42.58 & 37.56 & 61.20 & 46.55 & 85.47 & 80.13 & 82.71 & 70.13 \\
 & Java  & 16.15 & 49.33 & 24.34 & 32.69 & 50.00 & 39.53   & 88.88 & 70.46 & 78.61 & 63.71\\
 & \cellcolor{gray!20}Overall  & \cellcolor{gray!20}28.31 & \cellcolor{gray!20}45.48 & \cellcolor{gray!20}34.90 & \cellcolor{gray!20}36.51 & \cellcolor{gray!20}58.66 & \cellcolor{gray!20}45.01   & \cellcolor{gray!20}86.69 & \cellcolor{gray!20}76.30 & \cellcolor{gray!20}81.16 & \cellcolor{gray!20}67.79\\

\midrule
\midrule

\multirow{3}{*}{Codellama-7B} &Python   & 97.68 & 18.31 & 30.84 & 55.02 & 19.29 & 28.57 & 0.84 & 63.63 & 1.67 & 10.89\\
 & Java   & 95.19 & 33.79 & 49.88 & 9.61 & 5.81 & 7.24 & 5.01 & 79.31 & 9.42 & 25.10\\
 & \cellcolor{gray!20}Overall & \cellcolor{gray!20}96.40 & \cellcolor{gray!20}23.87 & \cellcolor{gray!20}38.26 & \cellcolor{gray!20}45.22 & \cellcolor{gray!20}17.44 & \cellcolor{gray!20}25.17 &\cellcolor{gray!20}2.33 & \cellcolor{gray!20}75.00 & \cellcolor{gray!20}4.52  & \cellcolor{gray!20}16.07 \\

\midrule

\multirow{3}{*}{Deepseek-coder-6.7B} &Python   & 0.0 & 0.0 & 0.0 & 4.16 & 16.66 & 6.66 & 96.40 & 72.82 & 82.97 & 70.52\\
 & Java   & 0.0 & 0.0 & 0.0 & 0.0 & 0.0 & 0.0 & 94.36 & 65.68 & 77.45 & 62.61\\
 & \cellcolor{gray!20}Overall & \cellcolor{gray!20}0.0 & \cellcolor{gray!20}0.0 & \cellcolor{gray!20}0.0 & \cellcolor{gray!20}2.85 & \cellcolor{gray!20}9.09 & \cellcolor{gray!20}4.34 &\cellcolor{gray!20}95.71 & \cellcolor{gray!20}70.27 & \cellcolor{gray!20}81.04  & \cellcolor{gray!20}67.67 \\

\midrule

\multirow{3}{*}{Mamba-Codestral-7B} &Python   & 6.48 & 24.13 & 10.21 & 7.40 & 82.35 & 13.59 & 98.66 & 71.36 & 82.82 & 68.30 \\
 & Java   & 7.86 & 81.81 & 14.34 & 1.92 & 100.0 & 3.77 & 99.34 & 68.46 & 81.06 & 66.18\\
 & \cellcolor{gray!20}Overall & \cellcolor{gray!20}7.19 & \cellcolor{gray!20}40.00 & \cellcolor{gray!20}12.19 & \cellcolor{gray!20}6.22 & \cellcolor{gray!20}83.33 & \cellcolor{gray!20}11.58 &\cellcolor{gray!20}98.91 & \cellcolor{gray!20}70.29 & \cellcolor{gray!20}82.18  & \cellcolor{gray!20}67.53 \\

\midrule

\multirow{3}{*}{Qwen-2.5-coder-32B} & Python  & 17.59 & 27.74 & 21.53 & 80.95 & 74.63 & 77.66 & 91.89 & 81.61 & 86.45 & 72.05\\
 & Java  & 10.04 & 79.31 & 17.83 & 78.85 & 75.93 & 77.36 & 96.95 & 71.89 & 82.56 & 67.63\\
 & \cellcolor{gray!20}Overall  & \cellcolor{gray!20}13.71 & \cellcolor{gray!20}36.75 & \cellcolor{gray!20}19.97 & \cellcolor{gray!20}80.50 & \cellcolor{gray!20}74.90 & \cellcolor{gray!20}77.60 & \cellcolor{gray!20}93.70 & \cellcolor{gray!20}77.28 & \cellcolor{gray!20}84.97 & \cellcolor{gray!20}70.44\\
\midrule
\midrule

\multirow{3}{*}{DCE-LLM} & Python & 98.61 & 89.50 & 93.83 & 97.88 & 97.37 & 97.63 & 99.52 & 99.16 & 99.34 & 97.00 \\
 & Java & 89.08 & 100.0 & 94.23 & 88.46 & 97.87 & 92.93 & 100.00 & 98.92 & 99.46 & 95.36 \\
 & \cellcolor{gray!20}Overall & \cellcolor{gray!20}\textbf{93.71} & \cellcolor{gray!20} \textbf{94.34} & \cellcolor{gray!20}\textbf{94.02} & \cellcolor{gray!20}\textbf{95.85} & \cellcolor{gray!20}\textbf{97.47} & \cellcolor{gray!20}\textbf{96.65} & \cellcolor{gray!20}\textbf{99.69} & \cellcolor{gray!20}\textbf{99.07} & \cellcolor{gray!20}\textbf{99.38} & \cellcolor{gray!20}\textbf{96.40} \\

\bottomrule
\end{tabular}
\label{taba3}
\end{table*}

\section{Example Prompts}

\subsection{Base Prompt}

\begin{figure}[ht]
\begin{minipage}{\linewidth}
\small
\noindent\makebox[\linewidth]{\rule{\linewidth}{1pt}}
\colorbox{codebg}{
\parbox{\dimexpr\linewidth-2\fboxsep}{
\textbf{\textcolor{cyan}{Input:}} \\
You are an experienced programmer expert in detecting dead code such as unused code and unreachable code. Given the following code, respond in the following format: \\
\textcolor{orange}{Dead code:} Yes or No \\
For each dead code,\\
\textcolor{gray}{Line Number:} Line number\\
\textcolor{olive}{Type:} Unused or Unreachable\\
\textcolor{purple}{Explanation:} Explanation for prediction\\

Finally, fix the code with: \\
\textcolor{brown}{Fixed Code:} Fixed code \\

\textcolor{blue}{Code:} [Code in Figure 1] \\

\textbf{\textcolor{magenta}{Output:}} \\
\textcolor{orange}{Dead code:} \textcolor{red}{Yes} \\
\textcolor{gray}{Line Number:} \textcolor{red}{4} \\
\textcolor{olive}{Type:} Unused \\
\textcolor{purple}{Explanation:} The variable s3 is defined but never used in any subsequent code. Instead of using s3, the code mistakenly uses the literal string `s3'. \\

\textcolor{brown}{Fixed Code:} [Code in Figure 1 after fix line 11]
}}
\noindent\makebox[\linewidth]{\rule{\linewidth}{1.5pt}}
\end{minipage}
\end{figure}

\subsection{Prompt with Suspect Lines}
\begin{figure}[ht]
\begin{minipage}{\linewidth}
\small
\noindent\makebox[\linewidth]{\rule{\linewidth}{1pt}}
\colorbox{codebg}{
\parbox{\dimexpr\linewidth-2\fboxsep}{
\textbf{\textcolor{cyan}{Input:}} \\
You are an experienced programmer expert in detecting dead code such as unused code and unreachable code. Given the following code and \textcolor{red}{suspect lines}, respond in the following format: \\
\textcolor{orange}{Dead code:} Yes or No \\
For each dead code,\\
\textcolor{gray}{Line Number:} Line number\\
\textcolor{olive}{Type:} Unused or Unreachable\\
\textcolor{purple}{Explanation:} Explanation for prediction\\

Finally, fix the code with: \\
\textcolor{brown}{Fixed Code:} Fixed code \\

\textcolor{blue}{Code:} [Code in Figure 1] \\
\textcolor{darkblue}{Suspect Lines:} [Suspect lines from the pivot model] \\

\textbf{\textcolor{magenta}{Output:}} \\
\textcolor{orange}{Dead code:} \textcolor{red}{Yes} \\
\textcolor{gray}{Line Number:} \textcolor{red}{4} \\
\textcolor{olive}{Type:} Unused \\
\textcolor{purple}{Explanation:} The variable s3 is defined but never used in any subsequent code. Instead of using s3, the code mistakenly uses the literal string `s3'. \\

\textcolor{brown}{Fixed Code:} [Code in Figure 1 after fix line 11]
}}
\noindent\makebox[\linewidth]{\rule{\linewidth}{1.5pt}}
\end{minipage}
\end{figure}

\end{document}